\documentclass[runningheads]{llncs}
\usepackage{graphicx}
\usepackage{amsmath}
\usepackage{booktabs}
\usepackage[table,usenames,dvipsnames]{xcolor}
\usepackage{threeparttable}
\usepackage{vcell}
\usepackage{subfig}
\usepackage{multirow}
\usepackage{float}
\usepackage{listings}
\usepackage{hyperref}
\hypersetup{
    colorlinks=true,
    allcolors=blue
    }
\usepackage{cleveref}
\usepackage{preprint_header}

\AtBeginDocument{%
  \providecommand\BibTeX{{%
    \normalfont B\kern-0.5em{\scshape i\kern-0.25em b}\kern-0.8em\TeX}}}

\begin{document}
\title{A Benchmark of PDF Information Extraction Tools using a Multi-Task and Multi-Domain Evaluation Framework for Academic Documents}

\titlerunning{A Benchmark of PDF Information Extraction Tools}

\author{Norman Meuschke\inst{1}\textsuperscript{,[\href{https://orcid.org/0000-0003-4648-8198}{ORCID}]} \and
Apurva Jagdale\inst{2} \and
Timo Spinde\inst{1}\textsuperscript{,[\href{https://orcid.org/0000-0003-3471-4127}{ORCID}]} \and
\\ Jelena Mitrovi\'c\inst{2,3}\textsuperscript{,[\href{https://orcid.org/0000-0003-3220-8749}{ORCID}]} \and
Bela Gipp\inst{1}\textsuperscript{,[\href{https://orcid.org/0000-0001-6522-3019}{ORCID}]}}
\authorrunning{Meuschke et al.}

\institute{University of Göttingen, 37073 Göttingen, Germany\\ 
\email{\{meuschke, spinde, gipp\}@uni-goettingen.de} \and
University of Passau, 94032 Passau, Germany\\
\email{\{apurva.jagdale, jelena.mitrovic\}@uni-passau.de} \and
The Institute for Artificial Intelligence R\&D of Serbia, 21000 Novi Sad, Serbia}

\maketitle
\thispagestyle{firststyle}
\begin{abstract}
Extracting information from academic PDF documents is crucial for numerous indexing, retrieval, and analysis use cases. Choosing the best tool to extract specific content elements is difficult because many, technically diverse tools are available, but recent performance benchmarks are rare. Moreover, such benchmarks typically cover only a few content elements like header metadata or bibliographic references and use smaller datasets from specific academic disciplines. We provide a large and diverse evaluation framework that supports more extraction tasks than most related datasets. Our framework builds upon DocBank, a multi-domain dataset of 1.5M annotated content elements extracted from 500K pages of research papers on arXiv. Using the new framework, we benchmark ten freely available tools in extracting document metadata, bibliographic references, tables, and other content elements from academic PDF documents. GROBID achieves the best metadata and reference extraction results, followed by CERMINE and Science Parse. For table extraction, Adobe Extract outperforms other tools, even though the performance is much lower than for other content elements. All tools struggle to extract lists, footers, and equations. We conclude that more research on improving and combining tools is necessary to achieve satisfactory extraction quality for most content elements. Evaluation datasets and frameworks like the one we present support this line of research. We make our data and code publicly available to contribute toward this goal.

\end{abstract}

\section{Introduction} \label{sec.introduction}
The Portable Document Format (PDF) is the most prevalent encoding for academic documents. Extracting information from academic PDF documents is crucial for numerous indexing, retrieval, and analysis tasks. Document search, recommendation, summarization, classification, knowledge base construction, question answering, and bibliometric analysis are just a few examples \cite{docbank}. 

However, the format's technical design makes information extraction challenging. Adobe designed PDF as a platform-independent, fixed-layout format by extending the PostScript~\cite{postscript} page description language. PDF focuses on encoding a document's visual layout to ensure a consistent appearance of the document across software and hardware platforms but includes little structural and semantic information on document elements.

Numerous tools for information extraction (IE) from PDF documents have been presented since the format's inception in 1993. The development of such tools has been subject to a fast-paced technological evolution of extraction approaches from rule-based algorithms, over statistical machine learning (ML) to deep learning (DL) models (cf. \Cref{sec.related_work}). Finding the best tool to extract specific content elements from PDF documents is currently difficult because:
\begin{enumerate}
	\item Typically, tools only support extracting a subset of the content elements in academic documents, e.g., title, authors, paragraphs, in-text citations, captions, tables, figures, equations, or references.
	\item Many information extraction tools, e.g., 12 of 35 tools we considered for our study, are no longer maintained or have become obsolete.
	\item Prior evaluations of information extraction tools often consider only specific content elements or use domain-specific corpora, which makes their results difficult to compare. Moreover, the most recent comprehensive benchmarks of information extraction tools were published in 2015 for metadata\footnote{For example author(s), title, affiliation(s), address(es), email(s)} \cite{hy3}, 2017 for body text \cite{icecite}, and 2018 for references\footnote{Refers to extracting the components of bibliographic references, e.g., author(s), title, venue, editor(s), volume, issue, page range, year of publication, etc.} \cite{hy2}, respectively. These evaluations do not reflect the latest technological advances in the field.
\end{enumerate}

To alleviate this knowledge gap and facilitate finding the best tool to extract specific elements from academic PDF documents, we comprehensively evaluate ten state-of-the-art non-commercial tools that consider eleven content elements based on a dataset of 500K pages from arXiv documents covering multiple fields.

\begin{center}
    \textbf{Our code, data, and resources are publicly available at} \\
\url{http://pdf-benchmark.gipplab.org}
\end{center}

\section{Related Work} \label{sec.related_work}
This section presents approaches for information extraction from PDF (\Cref{sec.ie_for_pdf}), labeled datasets suitable for training and evaluating PDF information extraction approaches, and prior evaluations of IE tools (\Cref{sec.datasets_evaluations}).

\subsection{Information Extraction from PDF Documents} \label{sec.ie_for_pdf}

\begin{table}
	\caption {Publications on information extraction from PDF documents.}\label{tab.extraction_approaches} 
	\setlength{\tabcolsep}{3.5pt}
	\renewcommand{\arraystretch}{1.2}
	\begin{threeparttable}[htb]
\begin{tabular}{lllll} 
\toprule
\textbf{Publication}\tnote{1}   & \textbf{Year} & \textbf{Task}\tnote{2}          & \textbf{Method}                                                   & \textbf{Training Dataset}\tnote{3}                                                                                                                              \\ 
\midrule
\vcell{Palermo \cite{rule3}}   & \vcell{1999}  & \vcell{M, ToC}        & \vcell{Rules}                                                       & \vcell{100 documents}                                                                                                                                       \\[-\rowheight]
\printcelltop     & \printcelltop & \printcelltop             & \printcelltop                                                       & \printcelltop                                                                                                                                           \\
\vcell{Klink \cite{rule2}}     & \vcell{2000}  & \vcell{M}             & \vcell{Rules}                                                       & \vcell{979 pages}                                                                                                                                       \\[-\rowheight]
\printcelltop     & \printcelltop & \printcelltop             & \printcelltop                                                       & \printcelltop                                                                                                                                           \\
\vcell{Giuffrida \cite{rule4}} & \vcell{2000}  & \vcell{M}             & \vcell{Rules}                                                       & \vcell{1,000 documents}                                                                                                                                     \\[-\rowheight]
\printcelltop     & \printcelltop & \printcelltop             & \printcelltop                                                       & \printcelltop                                                                                                                                           \\
\vcell{Aiello \cite{rule5}}    & \vcell{2002}  & \vcell{RO, Title}         & \vcell{Rules}                                                       & \vcell{1,000 pages}                                                                                                                                     \\[-\rowheight]
\printcelltop     & \printcelltop & \printcelltop             & \printcelltop                                                       & \printcelltop                                                                                                                                           \\
\vcell{Mao \cite{rule6}}       & \vcell{2004}  & \vcell{M}             & \vcell{\begin{tabular}[b]{@{}l@{}}OCR,\\Rules\end{tabular}}         & \vcell{309 documents}                                                                                                                                       \\[-\rowheight]
\printcelltop     & \printcelltop & \printcelltop             & \printcelltop                                                       & \printcelltop                                                                                                                                           \\
\vcell{Peng \cite{ml6}}      & \vcell{2004}  & \vcell{M, R}       & \vcell{CRF}                                                         & \vcell{CORA (500 refs.)}                                                                                                                           \\[-\rowheight]
\printcelltop     & \printcelltop & \printcelltop             & \printcelltop                                                       & \printcelltop                                                                                                                                           \\
\vcell{Day \cite{rule8}}       & \vcell{2007}  & \vcell{M, R}       & \vcell{Template}                                                    & \vcell{160,000 citations}                                                                                                                               \\[-\rowheight]
\printcelltop     & \printcelltop & \printcelltop             & \printcelltop                                                       & \printcelltop                                                                                                                                           \\
\vcell{Hetzner \cite{ml1}}   & \vcell{2008}  & \vcell{R}              & \vcell{HMM}                                                         & \vcell{CORA (500 refs.)}                                                                                                                            \\[-\rowheight]
\printcelltop     & \printcelltop & \printcelltop             & \printcelltop                                                       & \printcelltop                                                                                                                                           \\
\vcell{Councill \cite{tab5}}  & \vcell{2008}  & \vcell{R}              & \vcell{CRF}                                                         & \vcell{\begin{tabular}[b]{@{}l@{}}CORA (200 refs.), CiteSeer (200 refs.)\end{tabular}}                                                         \\[-\rowheight]
\printcelltop     & \printcelltop & \printcelltop             & \printcelltop                                                       & \printcelltop                                                                                                                                           \\
\vcell{Lopez \cite{grobid1}}     & \vcell{2009}  & \vcell{B, M, R} & \vcell{CRF, DL}                                                     & \vcell{None}                                                                                                                                            \\[-\rowheight]
\printcelltop     & \printcelltop & \printcelltop             & \printcelltop                                                       & \printcelltop                                                                                                                                           \\
\vcell{Cui \cite{rule7}}       & \vcell{2010}  & \vcell{M}             & \vcell{HMM}                                                         & \vcell{400 documents}                                                                                                                                       \\[-\rowheight]
\printcelltop     & \printcelltop & \printcelltop             & \printcelltop                                                       & \printcelltop                                                                                                                                           \\
\vcell{Ojokoh \cite{ml5}}    & \vcell{2010}  & \vcell{M}             & \vcell{HMM}                                                         & \vcell{\begin{tabular}[b]{@{}l@{}}CORA (500 refs.), ManCreat\\FLUX-CiM (300 refs.),\end{tabular}}                                       \\[-\rowheight]
\printcelltop     & \printcelltop & \printcelltop             & \printcelltop                                                       & \printcelltop                                                                                                                                           \\
\vcell{Kern \cite{hy1}}      & \vcell{2012}  & \vcell{M}             & \vcell{HMM}                                                         & \vcell{\begin{tabular}[b]{@{}l@{}}E-prints, Mendeley, PubMed (19K entries)\end{tabular}}                                                         \\[-\rowheight]
\printcelltop     & \printcelltop & \printcelltop             & \printcelltop                                                       & \printcelltop                                                                                                                                           \\
\vcell{Bast \cite{icecite_system}}      & \vcell{2013}  & \vcell{B, M, R} & \vcell{Rules}                                                       & \vcell{\begin{tabular}[b]{@{}l@{}}DBLP (690 docs.), PubMed (500 docs.)\end{tabular}}                                                           \\[-\rowheight]
\printcelltop     & \printcelltop & \printcelltop             & \printcelltop                                                       & \printcelltop                                                                                                                                           \\
\vcell{Souza \cite{ml8}}     & \vcell{2014}  & \vcell{M}             & \vcell{CRF}                                                         & \vcell{100 documents}                                                                                                                                       \\[-\rowheight]
\printcelltop     & \printcelltop & \printcelltop             & \printcelltop                                                       & \printcelltop                                                                                                                                           \\
\vcell{Anzaroot \cite{tab6}}  & \vcell{2014}  & \vcell{R}              & \vcell{CRF}                                                         & \vcell{UMASS (1,800 refs.)}                                                                                                                             \\[-\rowheight]
\printcelltop     & \printcelltop & \printcelltop             & \printcelltop                                                       & \printcelltop                                                                                                                                           \\
\vcell{Vilnis \cite{tab7}}    & \vcell{2015}  & \vcell{R}              & \vcell{CRF}                                                         & \vcell{UMASS (1,800 refs.)}                                                                                                                             \\[-\rowheight]
\printcelltop     & \printcelltop & \printcelltop             & \printcelltop                                                       & \printcelltop                                                                                                                                           \\
\vcell{Tkaczyk \cite{hy3}}   & \vcell{2015}  & \vcell{B, M, R} & \vcell{\begin{tabular}[b]{@{}l@{}}CRF,\\Rules,\\SVM\end{tabular}}    & \vcell{\begin{tabular}[b]{@{}l@{}}CiteSeer (4,000 refs.), CORA (500 refs.),\\GROTOAP, PMC (53K docs.)\end{tabular}}  \\[-\rowheight]
\printcelltop     & \printcelltop & \printcelltop             & \printcelltop                                                       & \printcelltop                                                                                                                                           \\
\vcell{Bhardwaj \cite{CNN}}  & \vcell{2017}  & \vcell{R}              & \vcell{FCN}                                                         & \vcell{5,090 references}                                                                                                                                     \\[-\rowheight]
\printcelltop     & \printcelltop & \printcelltop             & \printcelltop                                                       & \printcelltop                                                                                                                                           \\
\vcell{Rodrigues \cite{bilstm}} & \vcell{2018}  & \vcell{R}              & \vcell{BiLSTM}                                                      & \vcell{40,000 references}                                                                                                                                    \\[-\rowheight]
\printcelltop     & \printcelltop & \printcelltop             & \printcelltop                                                       & \printcelltop                                                                                                                                           \\
\vcell{Prasad \cite{tab8}}    & \vcell{2018}  & \vcell{M, R}       & \vcell{CRF, DL}                                                     & \vcell{\begin{tabular}[b]{@{}l@{}}FLUX-CiM (300 refs.), CiteSeer (4,000 refs.)\end{tabular}}                                                  \\[-\rowheight]
\printcelltop     & \printcelltop & \printcelltop             & \printcelltop                                                       & \printcelltop                                                                                                                                           \\
\vcell{Jahongir \cite{bench3}}  & \vcell{2018}  & \vcell{M}             & \vcell{Rules}                                                       & \vcell{10,000 documents}                                                                                                                                    \\[-\rowheight]
\printcelltop     & \printcelltop & \printcelltop             & \printcelltop                                                       & \printcelltop                                                                                                                                           \\
\vcell{Torre \cite{bench5}}     & \vcell{2018}  & \vcell{B, M}       & \vcell{Rules}                                                       & \vcell{300 documents}                                                                                                                                       \\[-\rowheight]
\printcelltop     & \printcelltop & \printcelltop             & \printcelltop                                                       & \printcelltop                                                                                                                                           \\
\vcell{Rizvi \cite{tool3}}     & \vcell{2020}  & \vcell{R}              & \vcell{R-CNN}                                                       & \vcell{40,000 references}                                                                                                                                    \\[-\rowheight]
\printcelltop     & \printcelltop & \printcelltop             & \printcelltop                                                       & \printcelltop                                                                                                                                           \\
\vcell{Hashmi \cite{tool1}}    & \vcell{2020}  & \vcell{M}             & \vcell{Rules}                                                       & \vcell{45 documents}                                                                                                                                        \\[-\rowheight]
\printcelltop     & \printcelltop & \printcelltop             & \printcelltop                                                       & \printcelltop                                                                                                                                           \\
\vcell{Ahmed \cite{tool2}}     & \vcell{2020}  & \vcell{M}             & \vcell{Rules}                                                       & \vcell{150 documents}                                                                                                                                       \\[-\rowheight]
\printcelltop     & \printcelltop & \printcelltop             & \printcelltop                                                       & \printcelltop                                                                                                                                           \\
\vcell{Nikolaos \cite{2021robust}}  & \vcell{2021}  & \vcell{B, M, R} & \vcell{\begin{tabular}[b]{@{}l@{}}Attention,\\BiLSTM\end{tabular}} & \vcell{3,000 documents}                                                                                                                                     \\[-\rowheight]
\printcelltop     & \printcelltop & \printcelltop             & \printcelltop                                                       & \printcelltop                                                                                                                                           \\
\bottomrule
\end{tabular}
		\begin{tablenotes}
			\item [1] Publications in chronological order; the labels indicate the first author only.
			\item [2] (B) Body text, (M) Metadata, (R) References, (RO) Reading order, (ToC) Table of contents 
                \item [3] Domain-specific datasets: Computer Science: CiteSeer~\cite{dataset10}, CORA~\cite{dataset9}, DBLP~\cite{DBLP}, FLUX-CiM~\cite{tab1,tab2}, ManCreat~\cite{ml5}; Health Science: PubMed~\cite{PubMed}, PMC~\cite{PMC}
		\end{tablenotes}
	\end{threeparttable}
\end{table}
\Cref{tab.extraction_approaches} summarizes publications on PDF information extraction since 1999. For each publication, the table shows the primary technological approach and the training dataset. Eighteen of 27 approaches (67\%) employ machine learning or deep learning (DL) techniques, and the remainder rule-based extraction (Rules). Early tools rely on manually coded rules \cite{rule3}. Second-generation tools use statistical machine learning, e.g., based on Hidden Markov Models (HMM) \cite{stats}, Conditional Random Fields (CRF) \cite{crf}, and maximum entropy \cite{stats2}. The most recent information extraction tools employ Transformer models \cite{transformer}.

A preference for---in theory---more flexible and adaptive machine learning and deep learning techniques over case-specific rule-based algorithms is observable in \Cref{tab.extraction_approaches}. However, many training datasets are domain-specific, e.g., they exclusively consist of documents from Computer Science or Health Science, and comprise fewer than 500 documents. These two factors put the generalizability of the respective IE approaches into question. Notable exceptions like Ojokoh et al. \cite{ml5}, Kern et al. \cite{hy1}, and Tkaczyk et al. \cite{hy3} use multiple datasets covering different domains for training and evaluation. However, these approaches address specific tasks, i.e, header metadata extraction, reference extraction, or both.

Moreover, a literature survey by Mao et al. shows that most approaches for text extraction from PDF do not specify the ground-truth data and performance metrics they use, which impedes performance comparisons \cite{litsur}. A positive exception is a publication by Bast et al. \cite{icecite_system}, which presents a comprehensive evaluation framework for text extraction from PDF that includes a fine-grained specification of the performance measures used.

\subsection{Labeled Datasets and Prior Benchmarks} \label{sec.datasets_evaluations}
\Cref{tab.labeled_datasets} summarizes datasets usable for training and evaluating PDF information extraction approaches grouped by the type of ground-truth labels they offer. Most datasets exclusively offer labels for document metadata, references, or both.

\begin{table}[htb]
	\caption{Labeled datasets for information extraction from PDF documents.} \label{tab.labeled_datasets}
	\centering
	\renewcommand{\tabcolsep}{3.5pt}
		\begin{threeparttable}[t]
			\begin{tabular}{llr} 
\toprule
\textbf{Publication}\tnote{1} & \textbf{Size}        & \textbf{Ground-truth Labels}                                                                                                                         \\ 
\midrule
\vcell{Fan \cite{dataset1}}           & \vcell{147 documents}    & \vcell{Metadata}                                                                                                                                     \\[-\rowheight]
\printcelltop         & \printcelltop        & \printcelltop                                                                                                                                        \\
\vcell{F\"arber \cite{dataset2}}       & \vcell{90K documents}    & \vcell{References}                                                                                                                                   \\[-\rowheight]
\printcelltop         & \printcelltop        & \printcelltop                                                                                                                                        \\
\vcell{Grennan \cite{dataset5}}       & \vcell{1B references}     & \vcell{References}                                                                                                                                   \\[-\rowheight]
\printcelltop         & \printcelltop        & \printcelltop                                                                                                                                        \\
\vcell{Saier \cite{dataset3,dataset4}}         & \vcell{1M documents.}     & \vcell{References}                                                                                                                                   \\[-\rowheight]
\printcelltop         & \printcelltop        & \printcelltop                                                                                                                                        \\
\vcell{Ley \cite{dataset8,DBLP}}         & \vcell{6M documents}     & \vcell{Metadata, references}                                                                                                                         \\[-\rowheight]
\printcelltop         & \printcelltop        & \printcelltop                                                                                                                                        \\
\vcell{Mccallum \cite{dataset9}}      & \vcell{935 documents}    & \vcell{Metadata, references}                                                                                                                         \\[-\rowheight]
\printcelltop         & \printcelltop        & \printcelltop                                                                                                                                        \\
\vcell{Kyle \cite{dataset7}}          & \vcell{8.1M documents}   & \vcell{Metadata, references}                                                                                                                         \\[-\rowheight]
\printcelltop         & \printcelltop        & \printcelltop                                                                                                                                        \\
\vcell{Ororbia \cite{dataset10}}       & \vcell{6M documents.}     & \vcell{Metadata, references}                                                                                                                         \\[-\rowheight]
\printcelltop         & \printcelltop        & \printcelltop                                                                                                                                        \\
\vcell{Bast \cite{icecite}}          & \vcell{12,098 documents} & \vcell{Body text, sections, title}                                                                                                                   \\[-\rowheight]
\printcelltop         & \printcelltop        & \printcelltop                                                                                                                                        \\
\vcell{Li \cite{docbank}}            & \vcell{500K pages}   & \vcell{\begin{tabular}[b]{@{}r@{}}Captions, equations, figures, footers\\lists, metadata, paragraphs,\\references, sections, tables\end{tabular}}  \\[-\rowheight]
\printcelltop         & \printcelltop        & \printcelltop                                                                                                                                        \\
\bottomrule
\end{tabular}
			\begin{tablenotes}
                    \item [1] The labels indicate the first author only.
			\end{tablenotes}
		\end{threeparttable}
\end{table}

Only the DocBank dataset by Li et al. \cite{docbank} offers annotations for 12 diverse content elements in academic documents, including, figures, equations, tables, and captions. Most of these content elements have not been used for benchmark evaluations yet. DocBank is comparably large (500K pages from research papers published on arXiv in a four-year period). A downside of the DocBank dataset is its coarse-grained labels for references, which do not annotate the fields of bibliographic entries like the author, publisher, volume, or date, as do bibliography-specific datasets like unarXive \cite{dataset5} or S2ORC \cite{dataset7}.

\Cref{tab.prior_evaluations} shows PDF information extraction benchmarks performed since 1999. Few such works exist and were rarely repeated or updated, which is sub-optimal given that many tools receive updates frequently. Other tools become technologically obsolete or unmaintained. For instance, pdf-extract\footnote{\url{https://github.com/CrossRef/pdfextract}}, lapdftext\footnote{\url{https://github.com/BMKEG/lapdftext}},
PDFSSA4MET\footnote{\url{https://github.com/eliask/pdfssa4met}}, and
PDFMeat\footnote{\url{https://github.com/dimatura/pdfmeat}} are no longer maintained actively, while ParsCit\footnote{\url{https://github.com/knmnyn/ParsCit}} has been replaced by NeuralParsCit\footnote{\url{https://github.com/WING-NUS/Neural-ParsCit}} and SciWING\footnote{\url{https://github.com/abhinavkashyap/sciwing}}.

\begin{table}[htb]
	\caption{Benchmark evaluations of PDF information extraction approaches.}\label{tab.prior_evaluations}
	\centering
	\renewcommand{\tabcolsep}{5pt}
		\begin{threeparttable}[t]
\begin{tabular}{lllrr} 
\toprule
\textbf{Publication}\tnote{1} & \textbf{Dataset}                                                                                  & \textbf{Metrics}\tnote{2}       & \textbf{Tools} & \textbf{Labels}\tnote{3}                                                 \\ 
\midrule
\vcell{Granitzer \cite{bench1}}     & \vcell{\begin{tabular}[b]{@{}l@{}}E-prints (2,452 docs.),\\ Mendeley (20,672 docs.)\end{tabular}} & \vcell{$P$, $R$}        & \vcell{2}      & \vcell{M}                                                   \\[-\rowheight]
\printcelltop         & \printcelltop                                                                                     & \printcelltop           & \printcelltop  & \printcelltop                                                  \\
\vcell{Lipinski \cite{bela}}      & \vcell{arXiv (1,253 docs.)}                                                                       & \vcell{$Acc$}           & \vcell{7}      & \vcell{M}                                                   \\[-\rowheight]
\printcelltop         & \printcelltop                                                                                     & \printcelltop           & \printcelltop  & \printcelltop                                                  \\
\vcell{Bast \cite{icecite}}          & \vcell{arXiv (12,098 docs.)}                                                                      & \vcell{Custom}          & \vcell{14}     & \vcell{\begin{tabular}[b]{@{}r@{}}NL, Pa\\RO, W\end{tabular}}  \\[-\rowheight]
\printcelltop         & \printcelltop                                                                                     & \printcelltop           & \printcelltop  & \printcelltop                                                  \\
\vcell{K\"orner \cite{bench4}}       & \vcell{100 (German docs.)}                                                                        & \vcell{$P$, $R$, $F_1$} & \vcell{4}      & \vcell{Ref}                                                    \\[-\rowheight]
\printcelltop         & \printcelltop                                                                                     & \printcelltop           & \printcelltop  & \printcelltop                                                  \\
\vcell{Tkaczyk \cite{hy2}}       & \vcell{9,491 documents}                                                                           & \vcell{$P$, $R$, $F_1$} & \vcell{10}     & \vcell{Ref}                                                    \\[-\rowheight]
\printcelltop         & \printcelltop                                                                                     & \printcelltop           & \printcelltop  & \printcelltop                                                  \\
\vcell{Rizvi \cite{evaluationnew}}        & \vcell{8,766 references}                                                                          & \vcell{$F_1$}           & \vcell{4}      & \vcell{Ref}                                                    \\[-\rowheight]
\printcelltop         & \printcelltop                                                                                     & \printcelltop           & \printcelltop  & \printcelltop                                                  \\
\bottomrule
\end{tabular}			

            \begin{tablenotes}
                \item [1] The labels indicate the first author only.
                \item [2] ($P$) Precision, ($R$) Recall, ($F_1$) $F_1$-score, ($Acc$) Accuracy
                \item [3] (M) Metadata, (NL) New Line, (Pa) Paragraph, (Ref) Reference, (RO) Reading order, (W)~Words
            \end{tablenotes}
            \end{threeparttable}
\end{table}

As \Cref{tab.prior_evaluations} shows, the most extensive dataset used for evaluating PDF information extraction tools so far contains approx. 24,000 documents. This number is small compared to the sizes of datasets available for this task, shown in \Cref{tab.labeled_datasets}. Most studies focused on exclusively evaluating metadata and reference extraction (see also \Cref{tab.prior_evaluations}). An exception is a benchmark by Bast and Korzen \cite{icecite}, which evaluated spurious and missing words, paragraphs, and new lines for 14 tools but used a comparably small dataset of approx. 10K documents.

We conclude from our review of related work that (1) recent benchmarks of information extraction tools for PDF are rare, (2) mostly analyze metadata extraction, (3) use small, domain-specific datasets, and (4) include tools that have become obsolete or unmaintained. (5) A variety of suitably labeled datasets have not been used to evaluate information extraction tools for PDF documents yet. Therefore, we see the need for benchmarking state-of-the-art PDF information extraction tools on a large labeled dataset of academic documents covering multiple domains and containing diverse content elements.

\section{Methodology}
This section presents the experimental setup of our study by describing the tools we evaluate (\Cref{sec.eval_tools}), the dataset we use (\Cref{sec.eval_dataset}), and the procedure we follow (\Cref{sec.eval_procedure}).

\subsection{Evaluated Tools}\label{sec.eval_tools}
We chose ten actively maintained non-commercial open-source tools that we categorize by extraction tasks. 
\begin{enumerate}
	\item \textbf{Metadata Extraction} includes tools to extract titles, authors, abstracts, and similar document metadata.
	\item \textbf{Reference Extraction} comprises tools to access and parse bibliographic reference strings into fields like author names, publication titles, and venue.
	\item \textbf{Table Extraction} refers to tools that allow accessing both the structure and data of tables.
	\item \textbf{General Extraction} subsumes tools to extract, e.g., paragraphs, sections, figures, captions, equations, lists, or footers.
\end{enumerate}
For each of the tools we evaluate, Table \ref{tab.evaluated_tools} shows the version, supported extraction task(s), primary technological approach, and output format. Hereafter, we briefly describe each tool, focusing on its technological approach.

\begin{table}[!ht]
	\caption{Overview of evaluated information extraction tools.} \label{tab.evaluated_tools}
	\centering
	\renewcommand{\tabcolsep}{3.5pt}
	\resizebox{\columnwidth}{!}{
		\begin{threeparttable}[t]
			\begin{tabular}{llllr} 
\toprule
\textbf{Tool}         & \textbf{Version} & \textbf{Task}\tnote{1}  & \textbf{Technology}                 & \textbf{Output}         \\ 
\midrule
\vcell{Adobe Extract} & \vcell{1.0}      & \vcell{G, T}       & \vcell{Adobe Sensei AI Framework}   & \vcell{JSON, XLSX}      \\[-\rowheight]
\printcelltop         & \printcelltop    & \printcelltop      & \printcelltop                       & \printcelltop           \\
\vcell{Apache Tika}   & \vcell{2.0.0}    & \vcell{G}          & \vcell{Apache PDFBox}               & \vcell{TXT}             \\[-\rowheight]
\printcelltop         & \printcelltop    & \printcelltop      & \printcelltop                       & \printcelltop           \\
\vcell{Camelot}       & \vcell{0.10.1}   & \vcell{T}          & \vcell{OpenCV, PDFMiner~}           & \vcell{CSV, Dataframe}  \\[-\rowheight]
\printcelltop         & \printcelltop    & \printcelltop      & \printcelltop                       & \printcelltop           \\
\vcell{CERMINE}       & \vcell{1.13}     & \vcell{G,~M, R}    & \vcell{CRF, iText, Rules, SVM}      & \vcell{JATS}        \\[-\rowheight]
\printcelltop         & \printcelltop    & \printcelltop      & \printcelltop                       & \printcelltop           \\
\vcell{GROBID}        & \vcell{0.7.0}    & \vcell{G, M, R, T} & \vcell{CRF, Deep Learning, Pdfalto} & \vcell{TEI XML}         \\[-\rowheight]
\printcelltop         & \printcelltop    & \printcelltop      & \printcelltop                       & \printcelltop           \\
\vcell{PdfAct}        & \vcell{n/a}      & \vcell{G, M, R, T} & \vcell{pdftotext, rules}            & \vcell{JSON, TXT, XML}  \\[-\rowheight]
\printcelltop         & \printcelltop    & \printcelltop      & \printcelltop                       & \printcelltop           \\
\vcell{PyMuPDF}       & \vcell{1.19.1}   & \vcell{G}          & \vcell{OCR, tesseract}              & \vcell{TXT}             \\[-\rowheight]
\printcelltop         & \printcelltop    & \printcelltop      & \printcelltop                       & \printcelltop           \\
\vcell{RefExtract}    & \vcell{0.2.5}    & \vcell{R}          & \vcell{pdftotext, rules}            & \vcell{TXT}             \\[-\rowheight]
\printcelltop         & \printcelltop    & \printcelltop      & \printcelltop                       & \printcelltop           \\
\vcell{ScienceParse}  & \vcell{1.0}      & \vcell{G, M, R,}   & \vcell{CRF, pdffigures2, rules}     & \vcell{JSON}            \\[-\rowheight]
\printcelltop         & \printcelltop    & \printcelltop      & \printcelltop                       & \printcelltop           \\
\vcell{Tabula}        & \vcell{1.2.1}    & \vcell{T}          & \vcell{PDFBox, rules}               & \vcell{CSV, Dataframe}  \\[-\rowheight]
\printcelltop         & \printcelltop    & \printcelltop      & \printcelltop                       & \printcelltop           \\
\bottomrule
\end{tabular}
			\begin{tablenotes}
				\item[1] (G) General, (M) Metadata, (R) References, (T) Table 
			\end{tablenotes}
		\end{threeparttable}
	}
\end{table}

\textbf{Adobe Extract}\footnote{\url{https://www.adobe.io/apis/documentcloud/dcsdk/pdf-extract.html}} is a cloud-based API that allows extracting tables and numerous other content elements subsumed in the \textit{general extraction} category. The API employs the Adobe Sensei\footnote{\url{https://www.adobe.com/de/sensei.html}} AI and machine learning platform to understand the structure of PDF documents. To evaluate the Adobe Extract API, we used the Adobe PDFServices Python SDK\footnote{\url{https://github.com/adobe/pdfservices-python-sdk-samples}} to access the API's services.

\textbf{Apache Tika}\footnote{\url{https://tika.apache.org/}} allows metadata and content extraction in XML format. We used the tika-python\footnote{\url{https://github.com/chrismattmann/tika-python}} client to access the Tika REST API. Unfortunately, we found that tika-python only supports content (paragraphs) extraction.

\textbf{Camelot}\footnote{\url{https://github.com/camelot-dev/camelot}} can extract tables using either the \textit{Stream} or \textit{Lattice} modes. The former uses whitespace between cells and the latter table borders for table cell identification. For our experiments, we exclusively use the Stream mode, since our test documents are academic papers, in which tables typically use whitespace in favor of cell borders to delineate cells. The Stream mode internally utilizes the PDFMiner library\footnote{\url{https://github.com/pdfminer/pdfminer.six}} to extract characters that are subsequently grouped into words and sentences using whitespace margins. 

\textbf{CERMINE} \cite{hy3} offers metadata, reference, and general extraction capabilities. The tool employs the iText PDF toolkit\footnote{\url{https://github.com/itext}} for character extraction and the Docstrum\footnote{\url{https://github.com/chulwoopack/docstrum}} image segmentation algorithm for page segmentation of document images. CERMINE uses an SVM classifier implemented using the LibSVM\footnote{\url{https://github.com/cjlin1/libsvm}} library and rule-based algorithms for metadata extraction. For reference extraction, the tool employs $k$-means clustering, and Conditional Random Fields implemented using the MALLET\footnote{\url{http://mallet.cs.umass.edu/sequences.php}} toolkit for sequence labeling. CERMINE returns a single XML file containing the annotations for an entire PDF. We employ the Beautiful Soup\footnote{\url{https://www.crummy.com/software/BeautifulSoup/bs4/doc/}} library to filter CERMINE's output files for the annotations relevant to our evaluation.

\textbf{GROBID}\footnote{\url{https://github.com/kermitt2/grobid}} \cite{grobidimpl} supports all four extraction tasks. The tool allows using either feature-engineered CRF (default) or a combination of CRF and DL models realized using the DeLFT\footnote{\url{https://github.com/kermitt2/delft}} Deep Learning library, which is based on TensorFlow and Keras. GROBID uses a cascade of sequence labeling models for different components. The models in the model cascade use individual label sequencing algorithms and features; some models employ tokenizers. This approach offers flexibility by allowing model tuning and improves the model's maintainability. We evaluate the default CRF model with production settings (a recommended setting to improve the performance and availability of the GROBID server, according to the tool's documentation\footnote{\url{https://GROBID.readthedocs.io/en/latest/Troubleshooting/}}). 

\textbf{PdfAct} formerly called Icecite \cite{icecite_system} is a rule-based tool that supports all four extraction tasks, including the extraction of appendices, acknowledgments, and tables of contents. The tool uses the PDFBox\footnote{\url{http://pdfbox.apache.org/}} and pdftotext\footnote{\url{https://github.com/jalan/pdftotext}} PDF manipulation and content extraction libraries. We use the tool's JAR release\footnote{\url{https://github.com/ad-freiburg/pdfact}}.

\textbf{PyMuPDF}\footnote{\url{https://github.com/pymupdf/PyMuPDF}} extends the MuPDF\footnote{\url{https://mupdf.com/}} viewer library with font and image extraction, PDF joining, and file embedding. PyMuPDF uses tesseract\footnote{\url{https://github.com/tesseract-ocr/tesseract}} for OCR. PyMuPDF could not process files whose names include special characters.

\textbf{RefExtract}\footnote{\url{https://github.com/inspirehep/refextract}} is a reference extraction tool that uses pdftotext\footnote{\url{https://linux.die.net/man1/pdftotext}} and regular expressions. RefExtract returns annotations for the entire bibliography of a document. The ground-truth annotations in our dataset (cf. \Cref{sec.eval_dataset}), however, pertain to individual pages of documents and do not always cover the entire document. If ground-truth annotations are only available for a subset of the references in a document, we use regular expressions to filter RefExtract's output to those references with ground-truth labels.

\textbf{Science Parse}\footnote{\url{https://github.com/allenai/science-parse}} uses a CRF model trained on data from GROBID to extract the title, author, and references. It also employs a rule-based algorithm by Clark and Divvala \cite{pdftofig} to extract sections and paragraphs in JSON format.

\textbf{Tabula}\footnote{\url{https://github.com/chezou/tabula-py}} is a table extraction tool. Analogous to Camelot, Tabula offers a \textit{Stream} mode realized using PDFBox, and a \textit{Lattice} mode realized using OpenCV for table cell recognition. 

\subsection{Dataset} \label{sec.eval_dataset}

\begin{figure}[!ht]
	\centering
	\includegraphics[width=\linewidth,height=\linewidth,keepaspectratio]{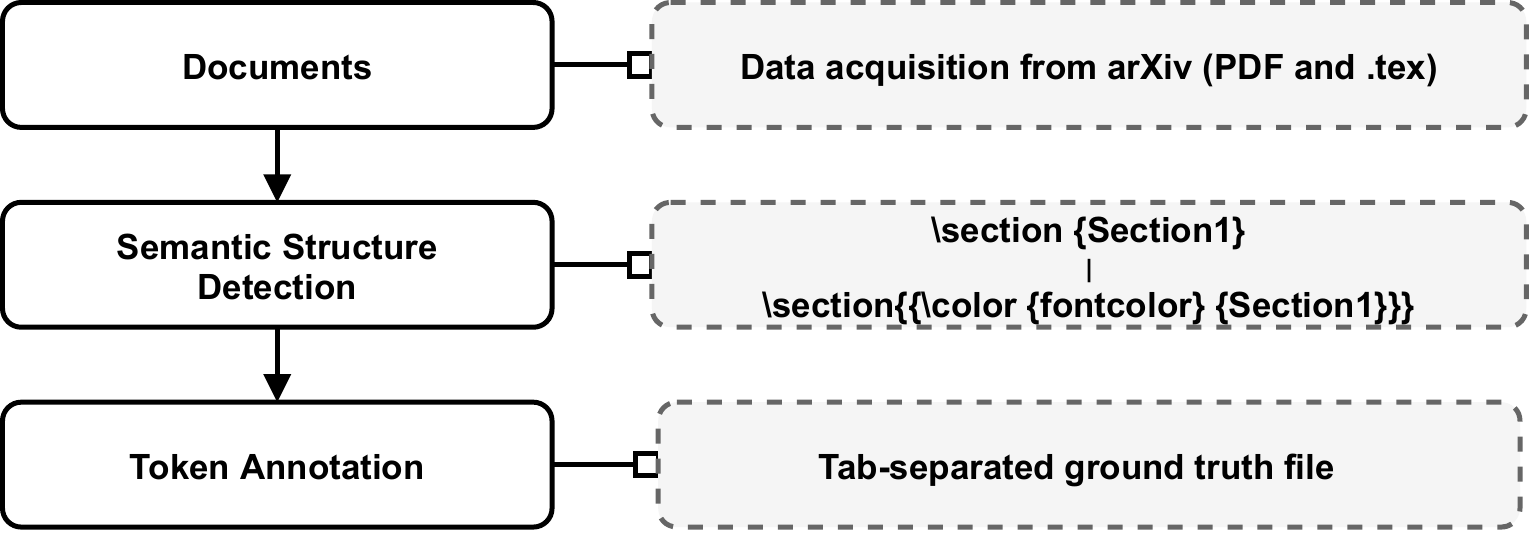}
	\caption{Process for generating the DocBank dataset.}
	\label{fig.docbank_generation}
\end{figure}
We use the DocBank\footnote{\url{https://github.com/doc-analysis/DocBank}} dataset, created by  Li et al. \cite{docbank}, for our experiments. Figure \ref{fig.docbank_generation} visualizes the process for compiling the dataset. First, the creators gathered arXiv documents, for which both the PDF and LaTeX source code was available. Li et al. then edited the LaTeX code to enable accurate automated annotations of content elements in the PDF version of the documents. For this purpose, they inserted commands that formatted content elements in specific colors. The center part of \Cref{fig.eval_procedure} shows the mapping of content elements to colors. In the last step, the dataset creators used PDFPlumber\footnote{\url{https://github.com/jsvine/pdfplumber}} and PDFMiner to extract and annotate relevant content elements by their color. DocBank provides the annotations as separate files for each document page in the dataset.

\begin{figure}[!ht]
	\centering
	\includegraphics[width=0.7\linewidth,height=\linewidth,keepaspectratio]{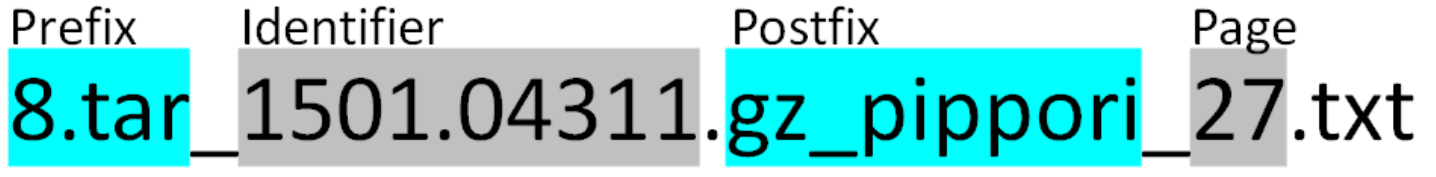}
	\caption{Naming scheme for DocBank's ground-truth files.} \label{fig.gt_file_naming}
\end{figure}

\Cref{tab.structure_gt_files} shows the structure of the tab-separated ground-truth files. Each line in the file refers to one component on the page and is structured as follows. Index 0 represents the token itself, e.g., a word. Indices 1-4 denote the bounding box information of the token, where (x0, y0) represents the top-left and (x1, y1) the bottom-right corner of the token in the PDF coordinate space. Indices 5-7 reflect the token's color in RGB notation, index 8 the token's font, and index 9 the label for the type of the content element. Each ground-truth file adheres to the naming scheme shown in \Cref{fig.gt_file_naming}.

\begin{table}[!ht]
	\caption{Structure of DocBank's plaintext ground-truth files.} \label{tab.structure_gt_files}
	\centering
        \renewcommand{\tabcolsep}{5pt}
		\begin{tabular}{ccccccccccc} 
			\toprule
			Index                       & 0                         & 1                      & 2                      & 3                      & 4                      & 5                     & 6                     & 7                     & 8                             & 9                          \\ 
			\hline
			\multicolumn{1}{l}{Content} & \multicolumn{1}{l}{token} & \multicolumn{1}{l}{x0} & \multicolumn{1}{l}{y0} & \multicolumn{1}{l}{x1} & \multicolumn{1}{l}{y1} & \multicolumn{1}{l}{R} & \multicolumn{1}{l}{G} & \multicolumn{1}{l}{B} & \multicolumn{1}{l}{font name} & \multicolumn{1}{l}{label}  \\
			\bottomrule
			\multicolumn{11}{c}{}\\
			\multicolumn{11}{c}{Source: \url{https://doc-analysis.github.io/docbank-page/index.html}.}
		\end{tabular}
\end{table}

The DocBank dataset offers ground-truth annotations for 1.5M content elements on 500K pages. Li et al. extracted the pages from arXiv papers in Physics, Mathematics, Computer Science, and numerous other fields published between 2014 and 2018. DocBank's large size, recency, diversity of included documents, number of annotated content elements, and high annotation quality due to the weakly supervised labeling approach make it an ideal choice for our purposes.

\subsection{Evaluation Procedure} \label{sec.eval_procedure} 

\begin{figure}[!ht]
	\centering
	\includegraphics[width=\linewidth,height=\linewidth,keepaspectratio]{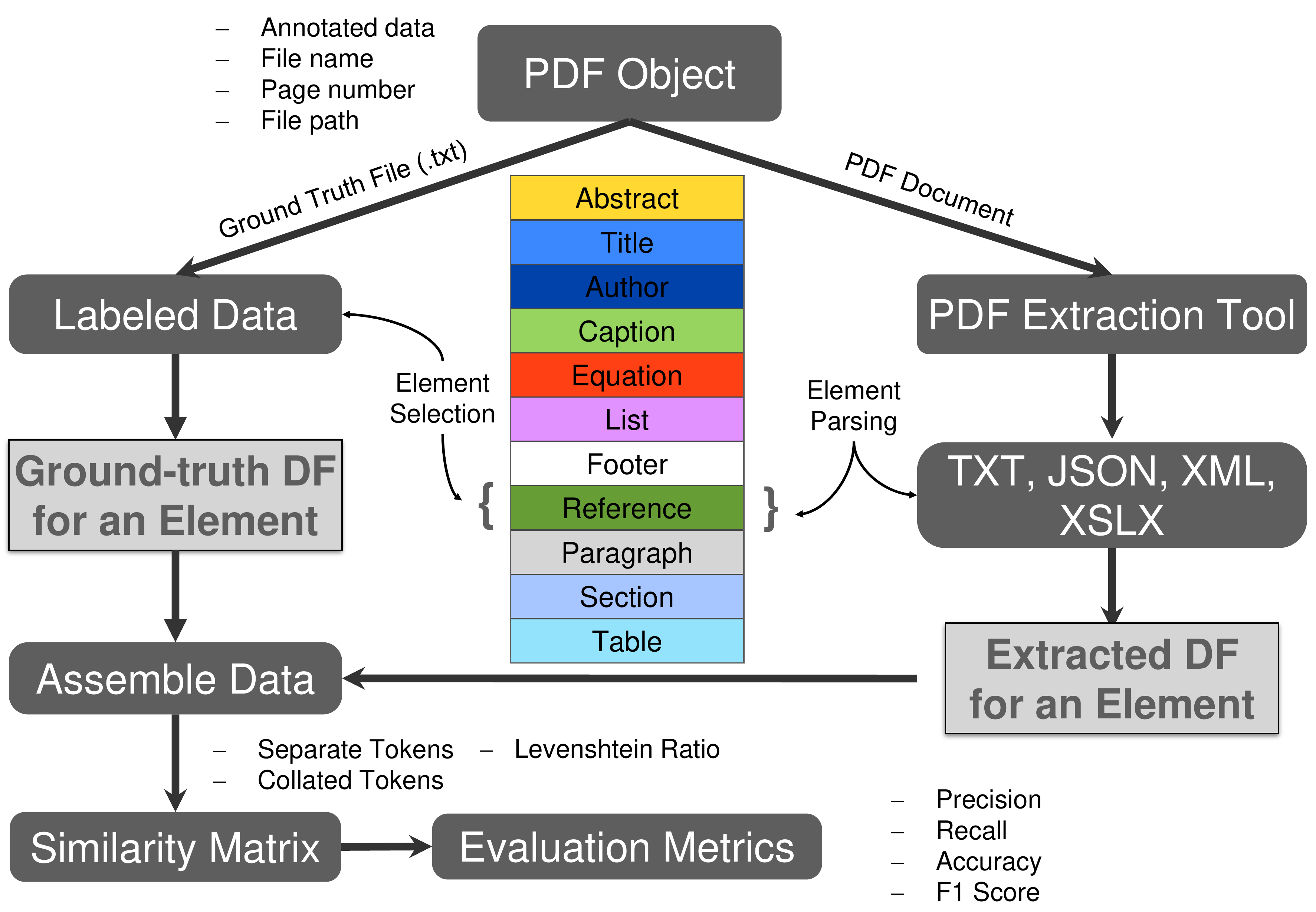}
	\caption{Overview of the procedure for comparing content elements extracted by IE tools to the ground-truth annotations and computing evaluation metrics.}
	\label{fig.eval_procedure}
\end{figure}

\Cref{fig.eval_procedure} shows our evaluation procedure. First, we select the PDF files whose associated ground-truth files contain relevant labels. For example, we search for ground-truth files containing \textit{reference} tokens to evaluate reference extraction tools. We include the PDF file, the ground-truth file, the document ID and page number obtainable from the file name (cf. \Cref{fig.gt_file_naming}), and the file path in a self-defined Python object (see \textit{PDF Object} in \Cref{fig.eval_procedure}). 

Then, the evaluation process splits into two branches whose goal is to create two pandas data frames---one holding the relevant ground-truth data, and the other the output of an information extraction tool. For this purpose, both the ground-truth files and the output files of IE tools are parsed and filtered for the relevant content elements. For example, to evaluate reference extraction via CERMINE, we exclusively parse reference tags from CERMINE's XML output file into a data frame (see \textit{Extracted DF} in \Cref{fig.eval_procedure}).

Finally, we convert both the \textit{ground-truth data frame} and the \textit{extracted data frame} into two formats for comparison and computing performance metrics. The first is the \textit{separate tokens} format, in which every token is represented as a row in the data frame. The second is the \textit{collated tokens} format, in which all tokens are combined into a single space-delimited row in the data frame. Separate tokens serve to compute a strict score for token-level extraction quality, whereas collated tokens yield a more lenient score intended to reflect a tool's average extraction quality for a class of content elements. We will explain the idea of both scores and their computation hereafter. 

We employ the \textit{Levenshtein Ratio} to quantify the similarity of extracted tokens and the ground-truth data for both the separate tokens and collated tokens format. \Cref{eq.lev_distance} defines the computation of the \textit{Levenshtein distance} of the extracted tokens $t_{e}$ and the ground-truth tokens $t_{g}$.

\begin{equation}\label{eq.lev_distance}
	\resizebox{\linewidth}{!}{$
		lev_{{t}_{e},{t}_{g}}(i,j)=
		\begin{cases} 
			max(i,j), & \text{if $min(i,j)=0$},\\ 
			min \begin{cases}
				lev_{{t}_{e},{t}_{g}}(i-1,j)+1\\
				lev_{{t}_{e},{t}_{g}}(i,j-1)+1\\
				lev_{{t}_{e},{t}_{g}}(i-1,j-1)+1_{({t}_{ei}\neq {t}_{ej})}
			    \end{cases}& \text{otherwise}.
		\end{cases}$}  
\end{equation}

\Cref{eq.lev_ratio} defines the derived Levenshtein Ratio score ($\gamma$). 

\begin{equation}\label{eq.lev_ratio}
	{\gamma\left( {t}_{e}, {t}_{g} \right) } = 1 -\frac{lev_{{t}_{e},{t}_{g}}(i,j)}{\left| {t}_{e} \right| + \left| {t}_{g} \right|}
\end{equation}

\Cref{eq.sim_matrix} shows the derivation of the \textit{similarity matrix} ($\Delta^{d}$) for a document ($d$), which contains the Levenshtein Ratio ($\gamma$) of every token in the extracted data frame with separate tokens $E^{s}$ of size $m$ and the ground-truth data frame with separate tokens $G^{s}$ of size $n$. 

\begin{equation}\label{eq.sim_matrix}
	{\Delta}_{m \times n}^{d} = {\gamma\left[ {E}_{i}^{s}, {G}_{j}^{s} \right] }_{i,j}^{m,n}
\end{equation}

Using the $m\times n$ similarity matrix, we compute the \textit{Precision} $P^d$ and \textit{Recall} $R^d$ scores according to \Cref{eq.precision} and \Cref{eq.recall}, respectively. As the numerator, we use the number of extracted tokens whose Levenshtein Ratio is larger or equal to 0.7. We chose this threshold for consistency with the experiments by Granitzer et al. \cite{bench1}. We then compute the $F_1^d$ score according to \Cref{eq.f1} as a token-level score for a tool's extraction quality. 

\begin{equation}\label{eq.precision}
	{P}^{d} = \frac{\# {\Delta}_{i,j}^{d} \ge 0.7}{m}
\end{equation}

\begin{equation}\label{eq.recall}
	{R}^{d} = \frac{\# {\Delta}_{i,j}^{d} \ge 0.7}{n}
\end{equation}

\begin{equation}\label{eq.f1}
	{F_1}^{d} = \frac{2 \times {P}^{d} \times {R}^{d}}{ {P}^{d} + {R}^{d}}
\end{equation}

Moreover, we compute the \textit{Accuracy} score $A^d$ reflecting a tool's average extraction quality for a class of tokens. To obtain $A^d$, we compute the Levenshtein Ratio $\gamma$ of the extracted tokens $E^{c}$ and ground-truth tokens $G^{c}$ in the collated tokens format, according to \Cref{eq.accuracy}.

\begin{equation}\label{eq.accuracy}
	{A}^{d} = {\gamma\left[ {E}^{c}, {G}^{c} \right] }
\end{equation}

\begin{figure}[!ht]
	\centering
	\includegraphics[width=\linewidth,height=\linewidth,keepaspectratio]{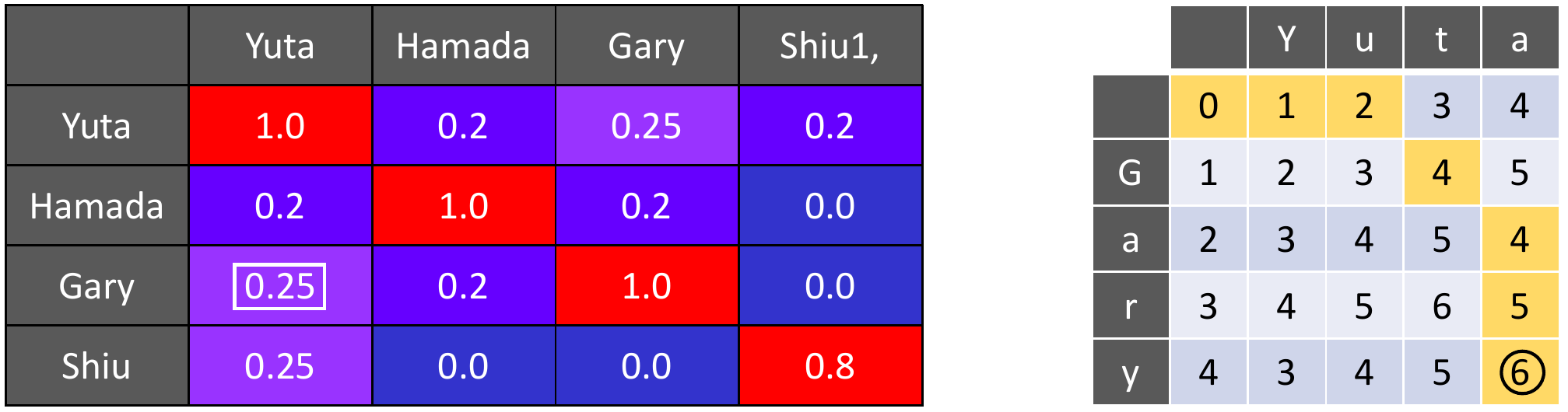}
	\caption{\underline{Left}: Similarity matrix for author names using separate tokens. \\ \underline{Right}: Computation of the Levenshtein distance (6) and the optimal edit transcript (yellow highlights) for two author names using dynamic programming.}
	\label{fig.sim_matrix_separate_tokens}
\end{figure}

\begin{figure}[!ht]
	\centering
	\includegraphics[width=0.8\linewidth,height=\linewidth,keepaspectratio]{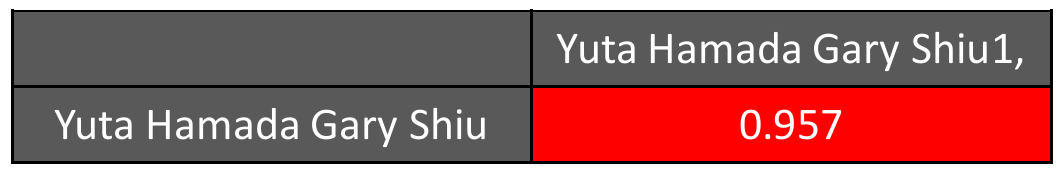}
	\caption{Similarity matrix for two sets of author names using collated tokens.}
	\label{fig.sim_matrix_collated_tokens}
\end{figure}

\Cref{fig.sim_matrix_separate_tokens} and \Cref{fig.sim_matrix_collated_tokens} show the similarity matrices for the author names 'Yuta,' 'Hamada,' 'Gary,' and 'Shiu' using separate and collated tokens, respectively. \Cref{fig.sim_matrix_separate_tokens} additionally shows an example computation of the Levenshtein Ratio for the strings \textit{Gary} and \textit{Yuta}. The strings have a Levenshtein distance of six and a cumulative string length of eight, which results in a Levenshtein Ratio of 0.25 that is entered into the similarity matrix. Figure \ref{fig.sim_matrix_collated_tokens} analogously exemplifies computing the Accuracy score of the two strings using collated tokens.

\newpage
\section{Results}
\begin{figure}[!htb]
	\centering
	\includegraphics[width=0.9\linewidth,height=\linewidth,keepaspectratio]{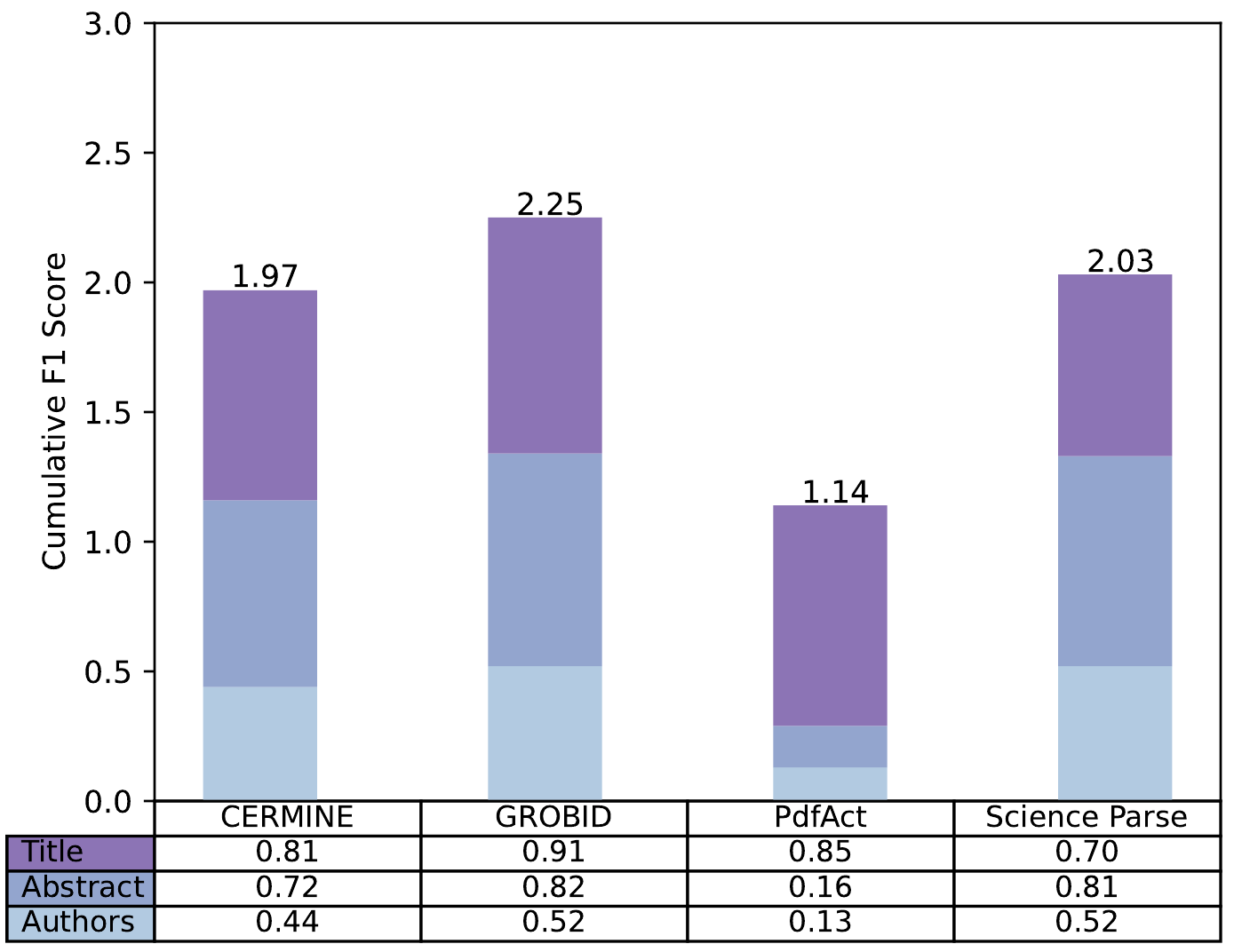}
	\caption{Results for metadata extraction.}
	\label{fig.results_metadata}
\end{figure}

\begin{figure}[!htb]
	\centering
	\includegraphics[width=0.9\linewidth,height=\linewidth,keepaspectratio]{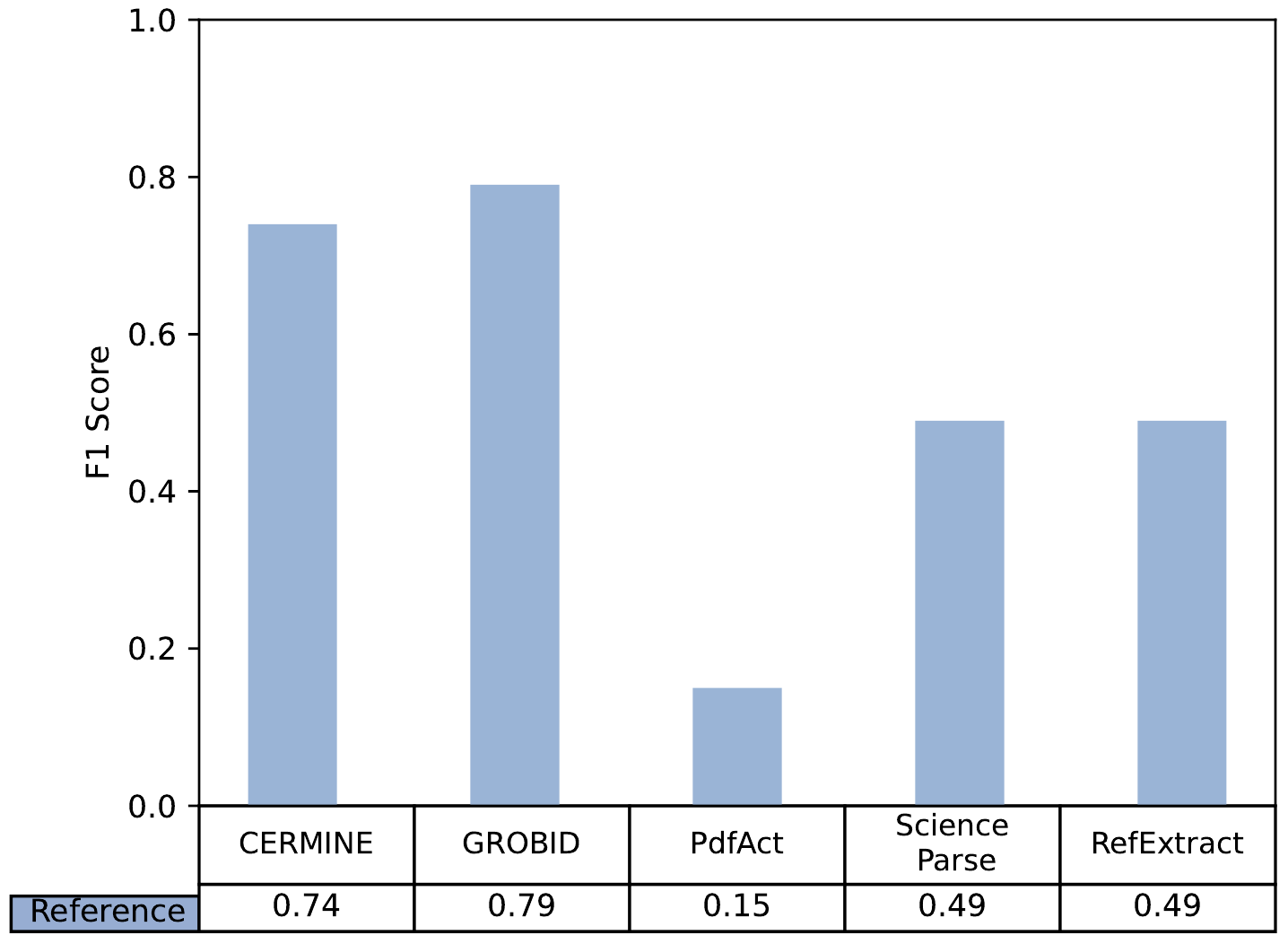}
	\caption{Results for reference extraction.}
	\label{fig.results_reference}
\end{figure}

\begin{figure}[!htb]
	\centering
	\includegraphics[width=0.9\linewidth,height=\linewidth,keepaspectratio]{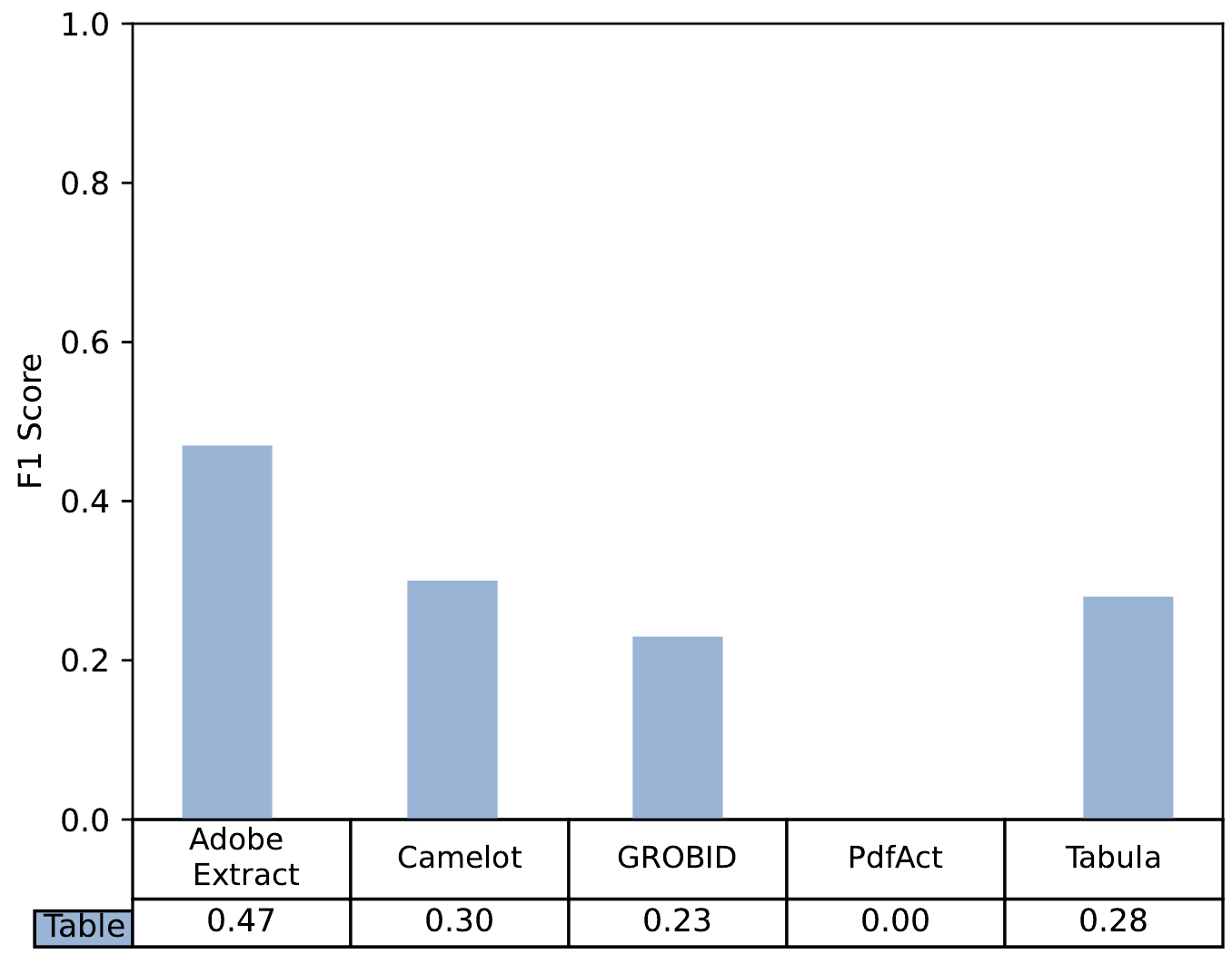}
	\caption{Results for table extraction.}
	\label{fig.results_table}
\end{figure}

\begin{figure}[!htb]
	\centering
	\includegraphics[width=0.9\linewidth,height=\linewidth,keepaspectratio]{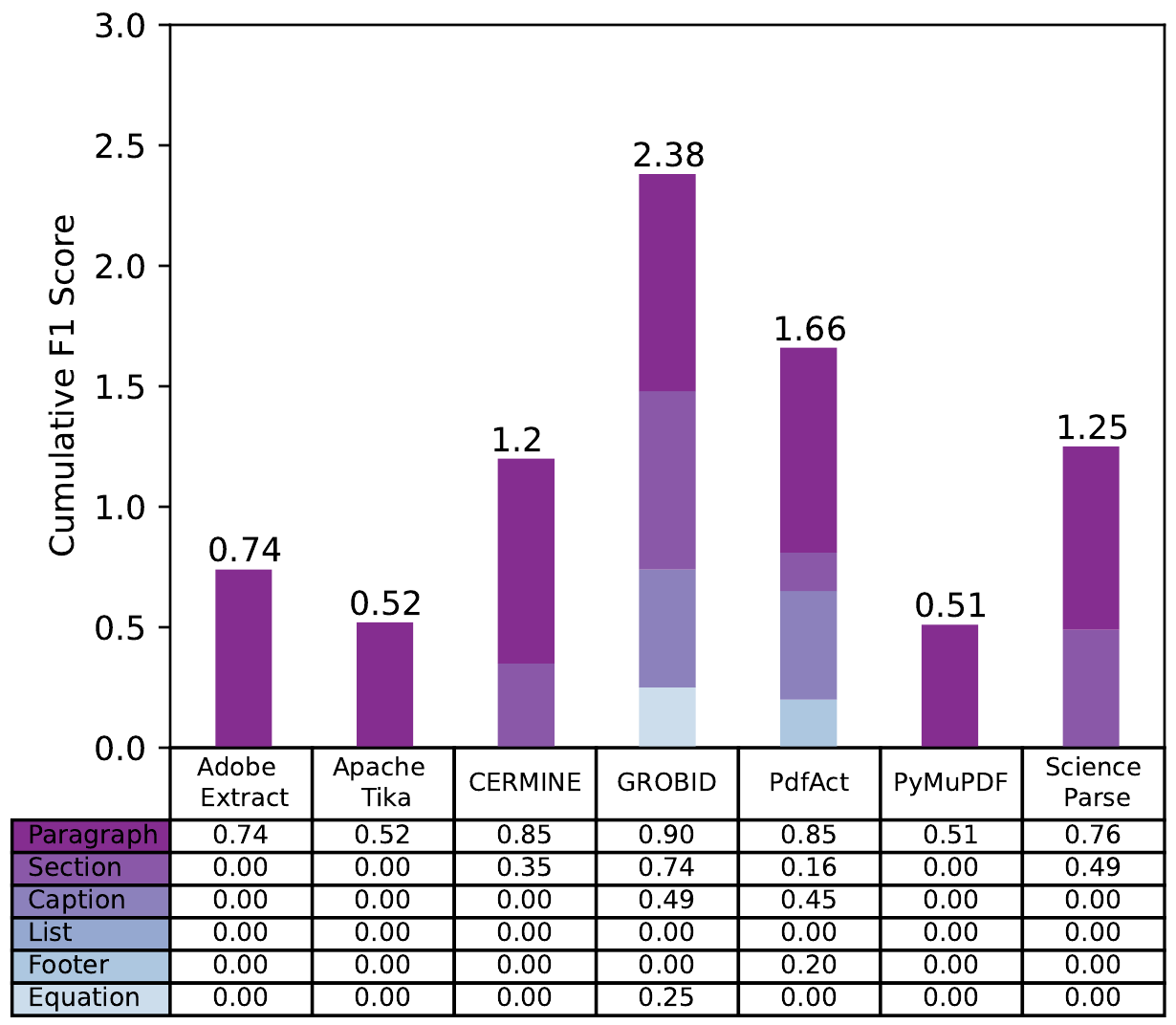}
	\caption{Results for general data extraction.}
	\label{fig.results_general}
\end{figure}

We present the evaluation results grouped by extraction task (see Figures \ref{fig.results_metadata}--\ref{fig.results_general}) and by tools (see \Cref{tab.results_by_tool}). This two-fold breakdown of the results facilitates identifying the best-performing tool for a specific extraction task or content element and allows for gauging the strengths and weaknesses of tools more easily. Note that the task-specific result visualizations (Figures \ref{fig.results_metadata}--\ref{fig.results_general}) only include tools that support the respective extraction task. See Table \ref{tab.evaluated_tools} for an overview of the evaluated tools and the extraction tasks they support.

\Cref{fig.results_metadata} shows the cumulative $F_1$ scores of CERMINE, GROBID, PdfAct, and Science Parse for the metadata extraction task, i.e., extracting title, abstract, and authors. Consequently, the best possible cumulative $F_1$ score equals three. Overall, GROBID performs best, achieving a cumulative $F_1$ score of 2.25 and individual $F_1$ scores of 0.91 for \textit{title}, 0.82 for \textit{abstract}, and 0.52 for \textit{authors}. Science Parse (2.03) and CERMINE (1.97) obtain comparable cumulative $F_1$ scores, while PdfAct has the lowest cumulative $F_1$ score of 1.14. However, PdfAct performs second-best for title extraction with a $F_1$ score of 0.85. The performance of all tools is worse for extracting authors than for titles and abstracts. It appears that machine-learning-based approaches like those of CERMINE, GROBID, and Science Parse perform better for metadata extraction than rule-based algorithms like the one implemented in PdfAct\footnote{See \Cref{tab.evaluated_tools} for more information on the tools' extraction approaches.}.

\Cref{fig.results_reference} shows the results for the reference extraction task. With a $F_1$ score of 0.79, GROBID also performs best for this task. CERMINE achieves the second rank with a $F_1$ score of 0.74, while Science Parse and RefExtract share the third rank with identical $F_1$ scores of 0.49. As for the metadata extraction task, PdfAct also achieves the lowest $F_1$ score of 0.15 for reference extraction. While both RefExtract and PdfAct employ pdftotext and regular expressions, GROBID performs efficient segregation of cascaded sequence labeling models\footnote{\url{https://grobid.readthedocs.io/en/latest/Principles/}} for diverse components, which can be the reason for its superior performance \cite{grobid1}.


\Cref{fig.results_table} depicts the results for the table extraction task. Adobe Extract outperforms the other tools with a $F_1$ score of 0.47. 
Camelot ($F_1=0.30$), Tabula ($F_1=0.28$), and GROBID ($F_1=0.23$) perform notably worse than Adobe Extract. Both Camelot and Tabula incorrectly treat two-column articles as tables and table captions as a part of the table region, which negatively affects their performance scores. The use of comparable \textit{Stream} and \textit{Lattice} modes in Camelot and Tabula (cf. \Cref{sec.eval_tools}) likely cause the tools' highly similar results. PdfAct did not produce an output for any of our test documents that contain tables, although the tool supposedly supports table extraction. The performance of all tools is significantly lower for table extraction than for other content elements, which is likely caused by the need to extract additional structural information. The difficulty of table extraction is also reflected by numerous issues that users opened on the matter in the GROBID GitHub repository\footnote{\url{https://github.com/kermitt2/grobid/issues/340}}.

\Cref{fig.results_general} visualizes the results for the general extraction task. GROBID achieves the highest cumulative $F_1$ score of 2.38, followed by PdfAct (cumulative $F_1=1.66$).
The cumulative $F_1$ scores of Science Parse (1.25), which only support paragraph and section extraction, and CERMINE (1.20) are much lower than GROBID's score and comparable to that of PdfAct. Apache Tika, PyMuPDF, and Adobe Extract can only extract paragraphs.

For paragraph extraction, GROBID (0.9), CERMINE (0.85), and PdfAct (0.85) obtained high $F_1$ scores with Science Parse (0.76) and Adobe Extract (0.74) following closely. Apache Tika (0.52) and PyMuPDF (0.51) achieved notably lower scores because the tools include other elements like sections, captions, lists, footers, and equations in paragraphs. 

Notably, only GROBID achieves a promising $F_1$ score of 0.74 for the extraction of sections. GROBID and PdfAct are the only tools that can partially extract captions. None of the tools is able to extract lists. Only PdfAct supports the extraction of footers but achieves a low $F_1$ score of 0.20. Only GROBID supports equation extraction but the extraction quality is comparatively low ($F_1=0.25$). To reduce the evaluation effort, we first tested the extraction of lists, footers, and equations on a two-months sample of the data covering January and February 2014. If a tool consistently obtained performance scores of 0, we did not continue with its evaluation. Following this procedure, we only evaluated GROBID and PdfAct on the full dataset. 

For the general extraction task, GROBID outperforms other tools due to its segmentation model\footnote{\url{https://grobid.readthedocs.io/en/latest/Principles/}}, which detects the main areas of documents based on layout features. Therefore, frequent content elements like paragraphs will not impact the extraction of rare elements from a non-body area by keeping the imbalanced classes in separate models. The cascading models used in GROBID also offer the flexibility to tune each model. Using layouts and structures as a basis for the process allows the association of simpler training data.

The breakdown of results by tools shown in \Cref{tab.results_by_tool} underscores the main takeaway point of the results' presentation for the individual extraction tasks. The tools' results differ greatly for different content elements. Certainly, no tool performs best for all elements, rather, even tools that perform well overall can fail completely for certain extraction tasks. The large amount of content elements whose extraction is either unsupported or only possible in poor quality indicates a large potential for improvement in future work. 

\begin{table}[H]
	\caption{Results grouped by extraction tool.} \label{tab.results_by_tool}
	\centering
	\renewcommand{\tabcolsep}{3.5pt}
	\begin{threeparttable}[t]
		\begin{tabular}{llllllll} 
\toprule
\textbf{Tool}\tnote{1}          & \textbf{Label} & \begin{tabular}[c]{@{}l@{}}\textbf{\# De-}\\\textbf{tected}\end{tabular} & \begin{tabular}[c]{@{}l@{}}\textbf{\# Pro-} \\ \textbf{cessed}\tnote{2}\end{tabular} & \textbf{$Acc$} & \textbf{$F_1$} & \textbf{$P$}  & \textbf{$R$}   \\ 
\midrule
\textbf{Adobe Extract} & Table          & 1,635                                                                 & 736                                                                      & \textbf{0.52}  & \textbf{0.47}  & \textbf{0.45} & \textbf{0.49}  \\
                       & Paragraph      & 3,985                                                                 & 3,088                                                                    & 0.85           & 0.74           & 0.72          & 0.76           \\ 
\midrule
\textbf{Apache Tika}   & Paragraph      & 339,603                                                               & 258,582                                                                  & 0.55           & 0.52           & 0.43          & 0.65           \\ 
\midrule
\textbf{Camelot}       & Table          & 16,289                                                                & \textbf{11,628}                                                          & 0.27           & 0.30           & 0.23          & 0.44           \\ 
\midrule
\textbf{CERMINE}       & Title          & 16,196                                                                & 14,501                                                                   & 0.84           & 0.81           & 0.81          & 0.81           \\
                       & Author         & \textbf{19,788}                                                       & 14,797                                                                   & 0.43           & 0.44           & 0.44          & 0.46           \\
                       & Abstract       & 19,342                                                                & 16,716                                                                   & 0.71           & 0.72           & 0.68          & 0.76           \\
                       & Reference      & \textbf{40,333}                                                       & 35,193                                                                   & 0.80           & 0.74           & 0.71          & 0.77           \\
                       & Paragraph      & 361,273                                                               & 348,160                                                                  & 0.89           & 0.85           & 0.83          & 0.87           \\
                       & Section        & \textbf{163,077}                                                      & 139,921                                                                  & 0.40           & 0.35           & 0.32          & 0.38           \\ 
\midrule
\textbf{GROBID}        & Title          & 16,196                                                                & 16,018                                                                   & \textbf{0.92}  & \textbf{0.91}  & \textbf{0.91} & \textbf{0.92}  \\
                       & Author         & \textbf{19,788}                                                       & \textbf{19,563}                                                          & \textbf{0.54}  & \textbf{0.52}  & \textbf{0.52} & \textbf{0.53}  \\
                       & Abstract       & 19,342                                                                & \textbf{18,714}                                                          & 0.82           & \textbf{0.82}  & \textbf{0.81} & 0.83           \\
                       & Reference      & \textbf{40,333}                                                       & \textbf{36,020}                                                          & \textbf{0.82}  & \textbf{0.79}  & \textbf{0.79} & \textbf{0.80}  \\
                       & Paragraph      & 361,273                                                               & \textbf{358,730}                                                         & \textbf{0.90}  & \textbf{0.90}  & \textbf{0.89} & \textbf{0.91}  \\
                       & Section        & \textbf{163,077}                                                      & \textbf{163,037}                                                         & \textbf{0.77}  & \textbf{0.74}  & \textbf{0.73} & \textbf{0.76}  \\
                       & Caption        & \textbf{90,606}                                                       & \textbf{62,445}                                                          & \textbf{0.57}  & \textbf{0.49}  & \textbf{0.47} & 0.51           \\
                       & Table          & \textbf{16,740}                                                       & 8,633                                                                    & 0.24           & 0.23           & 0.23          & 0.23           \\
                       & Equation       & \textbf{142,736}                                                      & \textbf{96,560}                                                          & \textbf{0.26}  & \textbf{0.25}  & \textbf{0.20} & \textbf{0.32}  \\ 
\midrule
\textbf{PdfAct}        & Title          & \textbf{17,670}                                                       & \textbf{16,834}                                                          & 0.85           & 0.85           & 0.85          & 0.86           \\
                       & Author         & 13,110                                                                & 2,187                                                                    & 0.14           & 0.13           & 0.12          & 0.18           \\
                       & Abstract       & \textbf{21,470}                                                       & 4,683                                                                    & 0.17           & 0.16           & 0.15          & 0.20           \\
                       & Reference      & 30,263                                                                & 12,705                                                                   & 0.19           & 0.15           & 0.17          & 0.20           \\
                       & Paragraph      & \textbf{361,318}                                                      & 357,905                                                                  & 0.85           & 0.85           & 0.80          & 0.89           \\
                       & Section        & 129,361                                                               & 87,605                                                                   & 0.21           & 0.16           & 0.12          & 0.25           \\
                       & Caption        & 83,435                                                                & 53,314                                                                   & 0.45           & 0.45           & 0.40          & \textbf{0.52}  \\
                       & Footer         & \textbf{32,457}                                                       & \textbf{26,252}                                                          & \textbf{0.23}  & \textbf{0.20}  & \textbf{0.25} & \textbf{0.16}  \\ 
\midrule
\textbf{PyMuPDF}       & Paragraph      & 339,650                                                               & 258,383                                                                  & 0.55           & 0.51           & 0.41          & 0.65           \\ 
\midrule
\textbf{RefExtract}    & Reference      & \textbf{40,333}                                                       & 38,405                                                                   & 0.55           & 0.49           & 0.44          & 0.55           \\ 
\midrule
\textbf{Science Parse} & Title          & 11,696                                                                & 11,687                                                                   & 0.79           & 0.70           & 0.70          & 0.70           \\
                       & Author         & 471                                                                   & 471                                                                      & \textbf{0.54}  & \textbf{0.52}  & \textbf{0.52} & \textbf{0.53}  \\
                       & Abstract       & 14,150                                                                & 14,149                                                                   & \textbf{0.83}  & 0.81           & 0.73          & \textbf{0.90}  \\
                       & Reference      & \textbf{40,333}                                                       & 35,200                                                                   & 0.55           & 0.49           & 0.49          & 0.50           \\
                       & Paragraph      & \textbf{361,318}                                                      & 355,529                                                                  & 0.79           & 0.76           & 0.76          & 0.76           \\
                       & Section        & \textbf{163,077}                                                      & 158,556                                                                  & 0.54           & 0.49           & 0.49          & 0.50           \\ 
\midrule
\textbf{Tabula}        & Table          & 10,361                                                                & 9,456                                                                    & 0.29           & 0.28           & 0.20          & 0.46           \\
\bottomrule
\end{tabular}
		\begin{tablenotes}
		\item [1] Boldface indicates the best value for each content element type.	
            \item [2] The differences in the number of detected and processed items are due to PDF Read Exceptions or Warnings. We label an item as processed if it has a non-zero $F_1$ score.
		\end{tablenotes}
	\end{threeparttable}
\end{table}

\section{Conclusion and Future Work}
\label{sec.conclusion}
We present an open evaluation framework for information extraction from academic PDF documents. Our framework uses the DocBank dataset \cite{docbank} offering 12 types and 1.5M annotated instances of content elements contained in 500K pages of arXiv papers from multiple disciplines. The dataset is larger, more topically diverse, and supports more extraction tasks than most related datasets.

We use the newly developed framework to benchmark the performance of ten freely available tools in extracting document metadata, bibliographic references, tables, and other content elements in academic PDF documents. GROBID, followed by CERMINE and Science Parse achieves the best results for the metadata and reference extraction tasks. For table extraction, Adobe Extract outperforms other tools, even though the performance is much lower than for other content elements. All tools struggle to extract lists, footers, and equations.

While DocBank covers more disciplines than other datasets, we see further diversification of the collection in terms of disciplines, document types, and content elements as a valuable task for future research. \Cref{tab.labeled_datasets} shows that more datasets suitable for information extraction from PDF documents are available but unused thus far. The weakly supervised annotation approach used for creating the DocBank dataset is transferable to other LaTeX document collections.

Apart from the dataset, our framework can incorporate additional tools and allows easy replacement of tools in case of updates. We intend to update and extend our performance benchmark in the future.

The extraction of tables, equations, footers, lists, and similar content elements poses the toughest challenge for tools in our benchmark. In recent work, Grennan et al.\cite{grennan2020synthetic} showed that the usage of synthetic datasets for model training can improve citation parsing. A similar approach could also be a promising direction for improving the access to currently hard-to-extract content elements.

Combining extraction approaches could lead to a one-fits-all extraction tool, which we consider desirable. The Sciencebeam-pipelines\footnote{\url{https://github.com/elifesciences/sciencebeam-pipelines}} project currently undertakes initial steps toward that goal. We hope that our evaluation framework will help to support this line of research by facilitating performance benchmarks of IE tools as part of a continuous development and integration process.

\bibliographystyle{splncs04}
\bibliography{bibliography.bib}
\end{document}